\documentclass[%
 reprint,
 amsmath,amssymb,
 aps,
]{revtex4-2}
\usepackage{color}
\usepackage{graphicx}
\usepackage{dcolumn}
\usepackage{bm}

\begin{document}

\preprint{APS/123-QED}

\title{Hubble tension and matter inhomogeneities: a theoretical perspective}

\author{San Martín, Marco}%
\email{mlsanmartin@uc.cl}
 \affiliation{Instituto de Astrofísica, Pontificia Universidad Católica de Chile, Avda. Vicuña Mackenna 4860, Santiago 7820436, Chile }


\author{Rubio, Carlos}
 \email{carlos.rubio@edu.uail.cl}
\affiliation{
 Facultad de Ingeniería y Ciencias, Universidad Adolfo Ibañez, Diagonal Las Torres 2640, Peñalolén, Santiago, Chile
}%

\date{\today}

\begin{abstract}
We have studied how local density perturbations could reconcile the Hubble tension. We reproduced a local void through a perturbed FLRW metric with a potential $\Phi$ which depends on both time and space. This method allowed us to obtain a perturbed luminosity distance, which is compared with both local and cosmological data. However, when constraining local cosmological parameters with previous results, we found that neither $\Lambda$CDM nor $\Lambda(\omega)$CDM cannot solve the Hubble tension.
\end{abstract}

\maketitle


\section{Introduction}

Knowledge of cosmology has great importance for the understanding of the physics laws that describe the universe. This is described with the standard model called $\Lambda$CDM, which indicates that the universe is mainly composed of radiation, baryonic matter (BM), dark matter (DM), and dark energy (DE) \citep[][]{Perlmutter_1999, Riess1998, Planck2018}.

This model indicates that approximately 5\% of the universe is made of baryons 69\% DE and 26\% DM, but these components are still not completely understood.
Despite $\Lambda$CDM has been adopted as the standard model to describe the universe at long scales, the improvement in the accuracy of the cosmological parameters has increased the tension with this model, for instance, problems with the curvature compatibility in the Planck data \citep{curvature} or the Hubble tension \citep{Riess_2016}.

In this work, we are interested in study the Hubble tension between the Planck and the local measurements of the Hubble constant ($H_0$). Regarding this parameter, the $H_0$ found by Planck is model-dependent because it requires all the perturbation theory associated with the model. In contrast, the local measurements \citep{Riess_2016, Riess_2018, Riess_2019, Riess_2021} only require a geometrical description of the distance that consider the expansion of the universe and how the flux of photons changes with the distance to the source. It is possible to delete the degeneracy between the absolute magnitude of the supernovae Ia (SNe-Ia) and the $H_0$ considering different calibrations, allowing to find an accurate $H_0$ value. 
Many observations has been constraining the $H_0$ values and claiming this tension with Planck. In general, the local measurements tends to be higher than the Planck $H_0$ value. For example, \cite{Sorce_2012} found a $H_0$ value of $75.2 \pm 3.0$ km Mpc$^{-1}$ s$^{-1}$ based on Sne-Ia. A few years later, \cite{Riess_2016} found an observed value $H_0 = 73.24\pm 1.74 $ km Mpc$^{-1}$ s$^{-1}$ using new parallaxes from Cepheids. This value is 3.4 $\sigma$ higher than $66.93\pm 0.62$ km Mpc$^{-1}$ s$^{-1}$ predicted by $\Lambda$CDM with Planck. But the discrepancy reduces to 2.1 $\sigma$ with respect to the prediction of $69.3\pm 0.7$ km Mpc$^{-1}$ s$^{-1}$ based on the comparably precise combination of WMAP+ACT+SPT+BAO observations. 

This tension between both observations has been widely discussed. For instance, there are different methods to calibrate the SNe-Ia. Regarding the cepheids calibration, the value of $H_0$ found in \citep{Riess_2018,Riess_2019,Breuval2020} is higher than $73$  km Mpc$^{-1}$ s$^{-1}$, where in the last one publication, \cite{Riess_2021} found that $H_0$ is $73.0 \pm 1.4$. All these values are far from the $H_0$ of Planck. However, the calibrations based on TRGB are lower than the cepheids measurements but higher than Planck, \citep{Yuan_2019, Soltis_2021} ( $72.4 \pm 2$ and $72.1 \pm 2.1$ km Mpc$^{-1}$ s$^{-1}$, respectively) and \citet{Freedman_2019,Freedman_2020} found that $H_0$ is $69.8 \pm 1.9$ and $69.6 \pm 1.9$ km Mpc$^{-1}$ s$^{-1}$, respectively. These values are lower than the previous based on cepheids but higher than the Planck value. \cite{Huang_2020} made a calibration based on Miras and found $H_0 = 73.30 \pm 4$ km Mpc$^{-1}$ s$^{-1}$, which is a high value compared with Planck, but also with a big uncertainty.

Many solutions have been proposed to explain this tension, such as extended models based on $\Lambda$CDM \citep{Guo_2019,Krishnan2021Htension}, time-varying DE density models  \citep{risaliti_lusso_2019}, or cosmography models \citep{Benetti_2019}. Others attempt modifications in the early-time physics \citep{vagnozzi2021consistency}, studying a component of dark radiation \citep{Bernal_2016} or analysing early physics related to the sound horizon \citep{Aylor_2019,krishnan2021does}. Many efforts related to recombination physics have been developed to solve the Hubble tension \citep{Agrawal2019, Lin_2019, Knox_2020}.

This controversy opens a window for new alternative theories based on modifications or variations of $\Lambda$CDM such as \citep{Haslbauer_2020,Camarena2018,Huang2016,Li_2013,Cede_o_2019,Xu_2019,Deser_2019,Anagnostopoulos_2019,PhysRevD.97.123504,SanMartin_2021}, other proposals introduce modifications in the physics of neutrinos \citep{Battye2014,Zhang_2014,Bernal_2016,Valentino_2016,Guo_2017_0,Feng2017,Zhao_2017,Guo_2017_1,Benetti_2017,Feng2018,Zhao2018,Benetti_2018,RoyChoudhury2019,Carneiro2019,Nakamura_2019,Vagnozzi_2020}, and others consider that DE can couple with DM: \citep{PhysRevD.88.023531,PhysRevD.89.103531,PhysRevD.96.123508,PhysRevD.96.043503,FENG2019100261,Yang_2018}.

Some independent studies support the idea that the tension is due more
to the physics rather than observational errors \citep{Benetti_2019, Bonvin_2016, Abbott_2018, Lemos_2019}. Others have found tension in the CMB analysis \citep{Addison_2016, curvature} or suggest errors in the values predicted by Planck CMB \citep{Spergel_2015}. Also, it has been suggested as a solution to include modifications in the Planck analysis through more free parameters and varying the Equation of State of DE \citep{DIVALENTINO2016242, PhysRevD.96.023523}. Also, it has been suggested that changing the cosmic sound horizon, the $H_0$ tension could be solved, but recently \citep{Jedamzik2021} has discarded this option because it develops tension with BAOs or galaxy weak lensing data.

In this work, we study the possibility that the local Hubble measurements differ from the Planck data because the local density of the universe is different from the global; in other words, the universe is not sufficiently homogeneous at the scales the SNe-Ia are observed. This idea has been previously studied by \cite{Bohringer,Haslbauer_2020,Keenan_2013,Shanks2018,10.1093/ptep/ptz066,Kenworthy_2019,PhysRevD.103.123539,Camarena_2022}. In general, they used the Lemaître-Tolman-Bondi (LTB) metric to describe the local void, or studied the influence of the subdensity on the velocities of the galaxies. In this paper, we developed a different approach: we assume a FLRW metric with scalar perturbations in a Newtonian gauge with no curvature to describe an inhomogeneity that is spherically symmetric.\\
This approach considers the temporal evolution of the field. This evolution is essential because from redshift 0.15 until now elapse approx. 2 Gyrs. This period allows the galaxies to move through space due to the field. In other words, our description allows a local spherically symmetric overdensity/subdensity, which affects the real dependence of the luminosity distance measured by a local observer centered on the origin of this inhomogeneity.\\This method considers a light ray traveling in an expanding universe until it reaches an observer located at the center of the local void. This void evolves over time according to linear perturbation equations. This void causes a perturbation to the observed redshift, which can be obtained from the perturbed geodesic presented in Chapter 3-B. Finally, to solve the perturbation equations, we consider Taylor expansions around the observer at the present time, located at the center of the void.\\
Finally, we compare our results with observational evidence of the local void, such as the amplitude of the local underdensity. In the results and conclusions, we comment on the differences between our method and others.

\section{Background cosmology}

In this work, we assume that the universe is flat and is composed of radiation, matter (BM and DM), and DE. The flat FLRW metric is given by

\begin{equation}
ds^2 = -c^2 dt^2 + a(t)^2 \left(dx^2+dy^2+dz^2\right).
\label{eq:FLRWmetric}
\end{equation}

Using the Einstein field equations 

\begin{equation}
G_{\mu \nu }=\kappa T_{\mu \nu },
\label{eq:einstein}
\end{equation}

where $\kappa \equiv \frac{8 \pi \mathbb{G}}{c^4}$, and the definition of a perfect fluid given by

\begin{equation}
T^{\mu\nu} = \left(\rho+p\right)u^\mu u^\nu+pg^{\mu\nu},
\label{eq:perfectfluid}
\end{equation}

where $\rho$ is the density and $p$ the pressure of the fluid, we get the Friedmann equation

\begin{equation}
H^2 \equiv \left(\frac{\dot{a}}{a}\right)^2  = \frac{8 \pi \mathbb{G} \rho}{3}.
\label{eq:friedmann}
\end{equation}

The density $\rho$ is determined by all the contributions ($m$ is for matter, which contains BM and DM):

\begin{equation}
\rho = \frac{3H_0^2}{8\pi \mathbb{G}} \left(\frac{\Omega_{r,0}}{a^4}+\frac{\Omega_{m,0}}{a^3}+\Omega_{\Lambda}\right).
\label{eq:density_total}
\end{equation}

Also, it is essential to define the density parameter $\Omega_{x,0}$, which is associated with the $x$ component of the universe as

\begin{equation}\label{def:density parameter}
\Omega_{x,0} = \frac{\rho_{x,0}}{\rho_{c,0}},
\end{equation}

where $\rho_{c}$ is the critical density of the universe, and the subscript $0$ denotes the parameter evaluated today. Then, the critical density today is $\rho_{c,0} = \frac{8\pi \mathbb{G}}{3 H_0^2}$, where $H_0$ is the Hubble constant. We emphasize that in standard cosmology, the DE is determined by a constant, but in this work, we work in a generalized case, where $\Omega_{\Lambda}$ depends on time (the scale factor is defined as $a_0 \equiv 1$ today).

From the Eq. \eqref{eq:friedmann}, the $\Lambda$CDM model preserves

\begin{equation}
\Omega_{r,0}+\Omega_{m,0}+\Omega_{\Lambda,0} = 1.
\label{eq:sum_omegas}
\end{equation}

In the following sections, we assume a time-dependent DE, a curvature parameter $k=0$, and two different Hubble constants $H_0$ denoted with different superscripts. The $H_0^\text{Pl}$ is an acronym for the \textit{Planck} measurements, and $H_0^\text{Lo}$ indicates \textit{Local} measurements. In our simplified description of the universe with inhomogeneities, the long distances measurements in cosmology contain information about the average universe, which is mainly flat and without inhomogeneities and anisotropies, as the cosmological principle says. In this scenario, the proper way to describe the universe is the \textit{background cosmology} given by the FLRW metric shown in Eq. \eqref{eq:FLRWmetric}. If there is a local inhomogeneity, this could be detectable depending on the amplitude of this perturbation. We assume a spherically symmetric inhomogeneity. This inhomogeneity changes the local physical description of the universe, and then, the parameters that an observer measures are incorrect, because he is assuming a non-perturbed metric. In this scenario, the Hubble constant measured by the observer who neglects the perturbative effect is given by $H_0^\text{Lo}$, and it is not the long-scale expansion rate. If researchers employ cosmographical models to represent cosmological parameters as series in time, scale factor, or redshift, and assume an unperturbed space-time, they may obtain values that are not consistent with the long-range cosmological parameters (such as the Planck values). The observable measured parameters, which are determined by assuming a non-perturbed FLRW metric, are denoted with the superscript \text{Lo} (from \textit{local}). In this context, there is an advantageous and common way of expanding the scale factor as a function of $\tau \equiv t-t_0$, $H_0$ (Hubble constant), $q_0$ (deceleration parameter) and $j_0$ (jerk). In our method, we utilize two distinct series with different interpretations. $a^\text{Pl}$ is employed to describe the unperturbed universe at large scales:

\begin{equation}
a^\text{Pl}(\tau) \approx 1 + H_0^\text{Pl} \tau -\frac{1}{2} q_0^\text{Pl}\left( H \tau\right)^2 + \frac{1}{6} j_0^\text{Pl} \left(H \tau\right)^3 + \mathcal{O}\left(\tau\right)^4.
\label{eq:a_series_Pl}
\end{equation}

By the other hand, the $a^\text{Lo}$ expansion describes the local universe, where the perturbations are important:

\begin{equation}
a^\text{Lo}(\tau) \approx 1 + H_0^\text{Lo} \tau -\frac{1}{2} q_0^\text{Lo}\left( H \tau\right)^2 + \frac{1}{6} j_0^\text{Lo} \left(H \tau\right)^3 + \mathcal{O}\left(\tau\right)^4.
\label{eq:a_series_Lo}
\end{equation}

The \textit{Local} and the \textit{Planck} parameters can coincide only when there is no perturbation. In another case, we have to study if the discrepancy can be explained assuming this perturbation. In the next section, we develop the effects of the perturbative terms in the luminosity distance. We emphasise that we consider two scenarios, the first is determined by $\Lambda$CDM with constant DE and a flat universe, and the second is described by $\Lambda(\omega)$CDM in a flat universe. In both cases the series expansions given by Eqs. \eqref{eq:a_series_Pl} and \eqref{eq:a_series_Lo} has to be consistent with the respective DE fluid equation.

\section{Perturbation theory}

We use a perturbed metric assuming the Newtonian Gauge with scalar perturbations represented by $\Psi(\tau,\vec{x})$ and $\Phi(\tau,\vec{x})$ in linear perturbation theory. Then, the most general metric is

\begin{eqnarray}
ds^2 &=& -c^2\left(1+2\Phi(\tau,\vec{x})\right)d\tau^2\nonumber\\ &+& a(\tau)^2\left(1-2\Psi(\tau,\vec{x})\right)\left(dx^2+dy^2+dz^2\right).
\label{eq:perturbed_metric}
\end{eqnarray}

As usual, the scalar perturbations are determined by the Bardeen potentials where the vector and tensor perturbations have been defined as $0$. $\Psi$ is related to the Newtonian potential, and $\Phi$ is associated with the spatial curvature perturbation.

Note that every perturbed quantity will be denoted with a super-index $^{(n)}$, which indicates the order of the perturbation.

We are interested in describing an expanding universe with a local inhomogeneity described by the Bardeen potentials. To describe the universe today, we can neglect the radiation contribution ($z < 0.15$ \cite{Riess_2016}). If the perturbations are caused by matter (BM and DM), then the first-order contribution from $T^{\mu\nu}$ can appear as a matter density perturbation, but not as a pressure perturbation. In other words, we are disregarding any interactions between matter, such as galaxy collisions or mergers. The DE appears as a constant density and pressure, but without perturbative terms. The perturbed matter term can move through space because it feels the Bardeen potentials which can evolve through time. This implies that the 4-velocity has to be perturbed, where we have included the DE term inside the energy-momentum tensor $\mathbf{T}_{\mu \nu } = T_{\mu \nu} - \Lambda g_{\mu \nu }$ and $\kappa = 8\pi \mathbb{G}/c^4$.

\subsection{Perturbed Einstein equations}

We expand the density, pressure and 4-velocity up to first order:

\begin{eqnarray}
\rho \approx \rho^{(0)}+\rho^{(1)},\\
p \approx p^{(0)}+p^{(1)},\\
u^\mu \approx {u^\mu}^{(0)}+{u^\mu}^{(1)}.
\end{eqnarray}

Using that $g_{\mu\nu}u^\mu u^\nu=-c^2$ and defining the velocity as $\left(v^x, v^y, v^z \right)$, then 

\begin{eqnarray}
u^\mu \approx (1-\Phi,v^x,v^y,v^z),\\
u_\mu \approx (-c^2(1+\Phi),a^2 v^x,a^2 v^y,a^2 v^z),
\end{eqnarray}

and with the Eq. \eqref{eq:perfectfluid} we get the perturbed Einstein equations ($\mathbf{G}^0{}_0{}^{(1)} = \kappa\mathbf{T}^0{}_0{}^{(1)}$, $\mathbf{G}^i{}_0{}^{(1)} = \kappa\mathbf{T}^i{}_0{}^{(1)}$, and $\mathbf{G}^i{}_i{}^{(1)} = \kappa\mathbf{T}^i{}_i{}^{(1)}$)

\begin{eqnarray}\label{eins1}
3H\dot{\Phi}-\frac{c^2}{a^2}\nabla^2 \Phi+3\Phi H^2 = -4\pi \mathbb{G}\rho_m^{(1)},\\
c^2\left(H\partial_i\Phi+\partial_i\dot{\Phi}\right) = -4\pi \mathbb{G} a^2\rho{}^{(0)}v^i\label{eins2},\\
\Phi\left(3H^2+2\dot{H}\right) +4H\dot{\Phi}+\ddot{\Phi} = 0.\label{timephi}
\end{eqnarray}
Note that  $\tau=t-t_0$ and $\dot{X}(\tau) \equiv dX(\tau)/dt$ where $X$ is an arbitrary function (this is an example). As we assume an isotropic fluid, then the components $i\neq j$ of Einstein field equations imply that $\Psi = \Phi$. Now, we introduce a velocity potential $v^N$ as $\vec{v} = -\nabla v^N$, then the Eqs. \eqref{eins1}-\eqref{timephi} are

\begin{eqnarray}\label{feins1}
c^2\nabla^2 \Phi=4\pi \mathbb{G}a^2\rho_m^{(1)}-3Ha^2(H\Phi+\dot{\Phi}),\\\label{feins2}
c^2(H\Phi+\dot{\Phi}) = 4\pi \mathbb{G} a^2\rho{}^{(0)}v^N,\\
\Phi\left(3H^2+2\dot{H}\right) +4H\dot{\Phi}+\ddot{\Phi} = 0.\label{feins3}
\end{eqnarray}

\subsection{Redshift dependence}

Considering the isotropic condition, the perturbed metric in spherical coordinates is

\begin{equation}
    ds^2=-\left(1+2\Phi\right)c^2d\tau^2+\left(1-2\Phi\right)a^2dr^2+ \left(1-2\Phi\right)a^2r^2 d\Omega^2,
\label{eq:metric_spherical}
\end{equation}

where the field $\Phi$ is spherically symmetric, $|\Phi(\tau,r)| \ll 1$ and $d\Omega^2 = d\theta^2+\sin(\theta)^2 d \phi^2$.

To find an expression for the luminosity distance that considers the effects produced by the perturbation,  we have to get an expression for the redshift as a function of the perturbations. In order to get these expression, we followed the Chapter 7.1 of \citep{weinberg_book}. A light ray traveling toward the center of the Robertson-Walker coordinate system from the direction $\hat{n}$ will have a co-moving radial coordinate $r$ related to $\tau$ by $$0=-(1+2\Phi)c^2d\tau^2+(1-2\Phi)a^2dr^2\,,$$ or in other words,

\begin{equation}\label{dr/dtau}
\frac{dr}{d\tau}=-c\left(\frac{a^2(1-2\Phi)}{1+2\Phi}\right)^{-1/2}\approx -\frac{c}{a}\left(1+2\Phi\right)+\mathcal{O}(\Phi^2).
\end{equation}

If the photon was emitted at time $\tau_e$, the relevant first-order solution of Eq. \eqref{dr/dtau} is 

\begin{equation}\label{r(tau)}
r(\tau)=s(\tau)-2c\int_{\tau_e}^\tau\frac{d\tau'}{a(\tau')}\Phi\left(\tau,s(\tau')\hat{n}\right)\,, 
\end{equation}

where $s(\tau)$ is the zeroth order solution for the radial coordinate which has the value $r_e$ at $\tau=\tau_e$: $$s(\tau)=r_e-c\int_{\tau_e}^\tau\frac{d\tau'}{a(\tau')}=c\int_\tau^{\tau_0}\frac{d\tau'}{a(\tau')}\,.$$
In particular, if the ray reaches $r=0$ at time $\tau_0=0$, the Eq. \eqref{r(tau)} gives 

\begin{eqnarray}
0&=&s(0)-2c\int_{\tau_e}^{0}\frac{d\tau}{a(\tau)}\Phi\left(\tau,s(\tau)\hat{n}\right)\nonumber\\
&=&r_e-c\int_{\tau_e}^{0}\frac{d\tau}{a(\tau)}\left(2\Phi\left(\tau,s(\tau)\hat{n}\right)+1\right)\,.\label{0=re} 
\end{eqnarray}

A time interval $\delta \tau_e$ between the departure of successive light wave crests at the time $\tau_e$ of emitted photons produces a time interval $\delta \tau_0$ between the arrival of successive crests at $\tau_0=0$ given by the variation of Eq. \eqref{0=re}: 
\begin{eqnarray}
0&=&\frac{c\delta \tau_e}{a(\tau_e)}\left[1+2\Phi(\tau_e,r_e\hat{n})-2c\int_{\tau_e}^{0}\frac{d\tau}{a(\tau)}\left(\frac{\partial\Phi(\tau,r\hat{n})}{\partial r}\right)_{r=s(\tau)}\right]\nonumber\\&-&\delta \tau_e\partial_rv^N(\tau_e,r_e\hat{n})-\frac{c\delta \tau_0}{a(0)}\left[1+2\Phi(0,0)\right]\label{dtau/dtaue}
\end{eqnarray}
(The term on the right-hand side involving the velocity potential $v^N$ arises from the change with time of the radial coordinate $r_e$ of the light source in Eq. \eqref{0=re}. We don't consider the variation of the argument $s(\tau)\hat{n}$ in $\Phi$, because to zeroth order $r_e$ and $\tau_e$ are related in such a way that $s(\tau)=0$ for all $r_e$, so its variation with $r_e$ is of first order, and the effect of this variation on $\Phi$ would be 
of second order). The total rate of change of $\Phi(\tau,s(\tau)\hat{n})$ is 
\begin{eqnarray*}
\frac{d}{d\tau}\Phi\left(\tau,s(\tau)\hat{n}\right)&=&\left(\frac{\partial}{\partial \tau}\Phi(\tau,r\hat{n})\right)_{r=s(\tau)}\\&-&\frac{c}{a(\tau)}\left(\frac{\partial \Phi(\tau,r\hat{n})}{\partial r}\right)_{r=s(\tau)}\,. 
\end{eqnarray*}

So the last equation may be written
\begin{eqnarray}\label{dTdt}
0=\frac{c\delta \tau_e}{a(\tau_e)}\left[1+2\Phi(0,0)-2\int_{\tau_e}^{0}d\tau\left\{\frac{\partial}{\partial \tau}\Phi(\tau,r\hat{n})\right\}_{r=s(\tau)}\right]\nonumber\\-\delta \tau_e\partial_r v^N(\tau_e,r_e\hat{n})-\frac{c\delta \tau_0}{a(0)}\left[1+2\Phi(0,0)\right].\nonumber\\
\end{eqnarray}

This gives the ratio of the coordinate time interval between emitted and received crests, but what we want is the ratio of the proper time intervals given by

$$\delta \tau^{proper}_e=\sqrt{1+2\Phi(\tau_e,r_e)}\delta \tau_e\,$$

and

$$ \delta\tau^{proper}_0=\sqrt{1+2\Phi(0,0)}\delta \tau_0,$$

which up to first order give the ratio of the received and emitted frequencies:

\begin{eqnarray*}
\frac{\nu_0}{\nu_e}=\frac{\delta \tau^{proper}_e}{\delta \tau^{proper}_0}=\frac{a(\tau_e)}{a(0)}\left[1+\Phi(\tau_e,r_e\hat{n})-\Phi(0,0)\right.\\+\left.2\int_{\tau_e}^{0}\left\{\frac{\partial}{\partial \tau}\Phi(\tau,r\hat{n})\right\}_{r=s(\tau)}d\tau\right.\\+\left.\frac{a(\tau_e)}{c}\partial_rv^N(\tau_e,r_e\hat{n})\right]\,.
\end{eqnarray*} 

Now, we define the redshift as usual,
\begin{equation*}
z_e+\delta z_e \equiv \frac{\lambda_0-\lambda_e}{\lambda_e} = \frac{\delta \tau^{proper}_0}{\delta \tau^{proper}_e}-1 = \frac{\nu_e}{\nu_0}-1,
\end{equation*}

then

\begin{equation}\label{zunperturbed}
    z_e=\frac{a(0)}{a(\tau_e)}-1
\end{equation}

and

\begin{eqnarray}\label{zperturbed}
\delta z_e&=&\frac{a(0)}{a(\tau)}\left[-\Phi(\tau_e,r_e\hat{n})+\Phi(0,0)\right.\\&-&\left.2\int_{\tau_e}^{0}\left\{\frac{\partial}{\partial \tau}\Phi(\tau,r\hat{n})\right\}_{r=s(\tau)}d\tau-\frac{a(\tau_e)}{c}\partial_rv^N(\tau,r_e\hat{n})\right].\nonumber
\end{eqnarray}
\\ 
 We emphasise that the cosmic time range corresponding to $z\in (0,0.15)$ is $t\in (13.721,11.821)$ Gyrs (using $H_0 \approx 69.6$ Mpc/(km s) and $\Omega_{m,0}$ = 0.286 in a flat universe). This time range allows $\Phi$ to vary. $\Phi$ evolves with the time, but it is always spherically symmetric. The time evolution is only produced by the movements of the galaxies in the perturbed metric. We are neglecting anisotropic effects, merging processes, and any other non-ideal effect. This is a simple approach.

\subsection{Luminosity distance}

We followed the standard derivation. A source of light emits photons of energy $h\nu_e$ during a time $\delta \tau_e$ and the observer measures photons with energy $h\nu_0$ during a time $\delta \tau_0$. Then the observed power $P_0$ (luminosity) is

\begin{equation}
P_0 = \frac{h\nu_0}{\delta \tau_0}.
\end{equation}
  
 Using the Eq. \eqref{dTdt} it follows
 
 \begin{equation}
P_0 = \frac{h\nu_e}{\delta \tau_e} \frac{1}{(1+z_e+\delta z_e)^2} \sqrt{\frac{1+2\Phi(0,0)}{1+2\Phi(\tau_e,r_e\hat{n})}}.
\end{equation}

Finally, using the definition of the emitted power, then the flux $F$ measured on a sphere of radius $l$ is given by

\begin{equation}
F = \frac{P_0}{4\pi l^2},
\end{equation}

where $l$ is determined by a measured made by the observer at time $\tau_0=0$. Then, following the same steps as in the standard cosmology, where $d\tau=0$ implies that the distance today is given by the spatial part evaluated at $\tau_0=0$

\begin{equation}
l(r) = \int_0^{l(r)} dl = a(0) \int_0^r dr =a(0) r,
\end{equation}

then, in the perturbed metric the proper distance measured today is

\begin{equation}
l(r) = \int_0^{l(r)} dl = a(0)\int_0^r \sqrt{1-2\Phi(0,r)}dr.
\end{equation}

This is the first difference introduced by the potential $\Phi(\tau,r)$. Then, the flux measured by the observer is

\begin{eqnarray}
F = \frac{h\nu_e}{\delta \tau_e}\frac{1}{4\pi\left(a(0)\int_0^{r_e} \sqrt{1-2\Phi(0,r)}dr\right)^2} \nonumber\\\frac{1}{(1+z_e+\delta z_e)^2}\sqrt{\frac{1+2\Phi(0,0)}{1+2\Phi(\tau_e,r_e\hat{n})}}.
\end{eqnarray}
 
Finally, the luminosity distance is defined such that $F\equiv \frac{P_e}{4\pi d_L^2}$, then
\begin{eqnarray}
d_L(z_e,\delta z_e) &=& \left(\frac{1+2\Phi(\tau,r\hat{n})}{1+2\Phi(0,0)}\right)^{1/4}\nonumber\\&\times&(1+z_e+\delta z_e)\int_0^r \sqrt{1-2\Phi(0,r)}dr,
\end{eqnarray}
where we consider that $a(0)=1$. Note that if $\Phi(\tau,r)=0$, the standard definition is recovered. Furthermore, we can expand at first order in $\Phi$ to obtain the luminosity distance

\begin{eqnarray}\label{luminositydistance}
d_L(z_e,\delta z_e)&=&(1+z_e+\delta z_e)r\nonumber\\&+&(1+z_e)\left[\frac{1}{2}(\Phi(\tau,r\hat{n})-\Phi(0,0))r\right.\nonumber\\&-&\left.\int_0^r\Phi(0,r\hat{n})dr\right]+\mathcal{O}(\Phi^2)\,.
\end{eqnarray}
where $\delta z$ was defined in the previous section.

\section{Perturbative solution}\label{Sec4}

In this section, we explain in detail the procedure of the expansions that we used to solve the potential $\Phi(\tau,r)$. Inspecting Eq. \ref{feins3}, we will propose, as an ansatz, that the solution for $\Phi$ can be separated into separable solutions. It is very important to remark that every element in the expansions is of the order of the perturbation $\Phi$. So, in every step, we linearised our solutions.

\subsection{$F(\tau)$}

First, we separated the potential in temporal and radial parts, $F(\tau)$ and $G(r)$,

\begin{eqnarray}\label{Phi(tau,r)}
&&\Phi(\tau,r) \equiv F(\tau)G(r), \\
&& F(\tau) \equiv \Phi_0+f_1\tau+f_2\tau^2 +f_3\tau^3+f_4\tau^4+\mathcal{O}\left(\tau^5\right), \label{F(t)} \\
&&G(r) \equiv 1+g_2 r^2+g_4 r^4 +\mathcal{O}\left(r^6\right) \label{G(r)}.
\end{eqnarray}

 The units of the coefficients are $[f_i]=$ s$^{-i}$, $[g_i]=$ Mpc$^{-i}$ and $\Phi_0$ is dimensionless. The Eq. \eqref{G(r)} preserves spherical symmetry. More important, any function $G(r)$ that has this symmetry could be expanded as Eq. \eqref{G(r)}, so our election for $G(r)$ at this level is completely general.
For our decomposition of $\Phi(\tau,r)$, we can solve for $F(\tau)$ using Eq. \eqref{timephi}, 

\begin{eqnarray}\label{eq:F_diff_eq}
F(\tau)\left(3H^2(\tau)+2\dot{H}(\tau)\right) +4H(\tau)\dot{F}(\tau)+\ddot{F}(\tau) = 0\,.\nonumber\\
\end{eqnarray}

Replacing the Eqs. \eqref{eq:a_series_Pl} and \eqref{F(t)} in Eq. \eqref{eq:F_diff_eq}, we get the solution for the coefficients $f_2$, $f_3$ and $f_4$. The expression are shown in Appendix \ref{f2f3f4}.

\subsection{$\tau(r)$}

We need to know how the time is related with the space for photons. To do that, we consider the radial  trajectory given by Eq. (\ref{dr/dtau}).
In the next step, we expand the cosmological time as

\begin{eqnarray}
    \tau(r)&=&\tau_1\left(\frac{r}{c}\right)+\tau_2\left(\frac{r}{c}\right)^2+\tau_3\left(\frac{r}{c}\right)^3+\tau_4\left(\frac{r}{c}\right)^4\nonumber\\&+&\tau_5\left(\frac{r}{c}\right)^5+O(r)^6\,.
\label{eq:tau}
\end{eqnarray}

We used a Taylor expansion because the radial trajectories relate the radial component with time around $r=0$, which corresponds to $\tau=0$. The units of $\tau_i$ are $s^{-i+1}$. Now we used Eq. \eqref{dr/dtau} to find the coefficients $\tau_i$. We include them in the Appendix \ref{coefficients}.

\subsection{Luminosity distance expansion}

With $\tau(r)$ for photons, we can find the expression for $r(z)$, where $z$ is the redshift. This was included in Appendix \ref{r(z)}. Finally, using the luminosity distance defined in Eq. \eqref{luminositydistance} we get

\begin{equation}\label{dLperturbed}
    d_L(z)=\frac{cz}{H_0^\text{Pl}}\left(\mathcal{D}_L^0+\mathcal{D}_L^1z+\mathcal{D}_L^2z^2+O(z)^3\right)\,,
\end{equation}

with

\begin{eqnarray}\label{cos solution1}
\mathcal{D}_L^0&=&1+\Phi_0+\frac{f_1}{{H_0^\text{Pl}}}-\frac{4c^2\Phi_0g_2}{3{H_0^\text{Pl}}^2}-\frac{4c^2f_1g_2}{3{H_0^\text{Pl}}^3}\,,\\
\mathcal{D}_L^1&=&-\frac{3 \Phi_{0} q_0^\text{Pl}}{2} + \Phi_{0} - \frac{q_0^\text{Pl}}{2} + \frac{1}{2} - \frac{3 f_{1} q_0^\text{Pl}}{2 H_0^\text{Pl}} + \frac{f_{1}}{2H_0^\text{Pl}}\nonumber\\ &-& \frac{\Phi_{0} c^{2} g_{2}}{\left(H_0^\text{Pl}\right)^{2}}+\frac{6c^2\Phi_0g_2q_0^\text{Pl}}{{H_0^\text{Pl}}^2}+\frac{14c^2f_1g_2q_0^\text{Pl}}{3{H_0^\text{Pl}}^3}\nonumber\\&-&\frac{10c^2f_1g_2}{3{H_0^\text{Pl}}^3},\\
\mathcal{D}_L^2&=&- \frac{\Phi_{0} j_0^\text{Pl}}{2} + \frac{5 \Phi_{0} {q_0^\text{Pl}}^{2}}{2} -\frac{\Phi_0}{4} - \frac{j_0^\text{Pl}}{6} + \frac{{q_0^\text{Pl}}^{2}}{2}\nonumber\\ &+& \frac{q_0^\text{Pl}}{6} - \frac{1}{6} - \frac{2 f_{1} j_0^\text{Pl}}{3 H_0^\text{Pl}} + \frac{5 f_{1} {q_0^\text{Pl}}^{2}}{2 H_0^\text{Pl}} - \frac{7f_{1}}{12 H_0^\text{Pl}}\nonumber\\
&+&\frac{26c^2\Phi_0g_2j_0^\text{Pl}}{9{H_0^\text{Pl}}^2}-\frac{18c^2\Phi_0g_2{q_0^\text{Pl}}^{2}}{{H_0^\text{Pl}}^2}+ \frac{16c^2 \Phi_{0} g_{2} q_0^\text{Pl}}{9{H_0^\text{Pl}}^{2}}\nonumber\\&-& \frac{5c^2\Phi_{0}  g_{2}}{18 {H_0^\text{Pl}}^{2}} - \frac{31c^{2} f_{1} g_{2}}{9{H_0^\text{Pl}}^{3}}+\frac{20c^2f_1g_2j_0^\text{Pl}}{9{H_0^\text{Pl}}^{3}}-\frac{38c^2f_1g_2{q_0^\text{Pl}}^{2}}{3{H_0^\text{Pl}}^{3}}\nonumber\\
&+&\frac{58c^2f_1g_2q_0^\text{Pl}}{9{H_0^\text{Pl}}^{3}}-\frac{8c^4\Phi_0g_4}{3{H_0^\text{Pl}}^{4}}-\frac{8c^4f_1g_4}{3{H_0^\text{Pl}}^{5}}\,.\label{cos solution3}
\end{eqnarray}

This expansion allow us to compare with local expansion of luminosity distance used by \cite{Riess_2016}. The standard expansion for $d_L^\text{Std}$ is 

\begin{equation}\label{dLStandard}
    d_L^\text{Std}(z)=\frac{cz}{H_0^\text{Lo}}\left(\mathcal{D}_L^{0,\text{Std}}+\mathcal{D}_L^{1,\text{Std}}z+\mathcal{D}_L^{2,\text{Std}}z^2+O(z)^3\right)\,,
\end{equation}

with

\begin{eqnarray}\label{std solution1}
\mathcal{D}_L^{0,\text{Std}}&=&1\,,\\
\mathcal{D}_L^{1,\text{Std}}&=&-\frac{1}{2}(-1+q_0^\text{Lo})\,,\\
\mathcal{D}_L^{2,\text{Std}}&=&-\frac{1}{6}(1-q_0^\text{Lo}-3{q_0^\text{Lo}}^2+j_0^\text{Lo})\,.\label{std solution3}
\end{eqnarray}

Comparing Eqs. \eqref{cos solution1}-\eqref{cos solution3} with Eqs. \eqref{std solution1}-\eqref{std solution3} allow us to find the parameters $\Phi_0$, $f_1$ and $g_2$ as a function of $H_0^\text{Lo}$, $q_0^\text{Lo}$ and $j_0^\text{Lo}$.

\section{Matter Perturbation}

\subsection{Equation for the matter inhomogeneity}

Now that we know $\Phi(\tau,r)$ in terms of $H_0^\text{Lo}$, $q_0^\text{Lo}$ and $j_0^\text{Lo}$ (with $H_0^\text{Pl}$, $q_0^\text{Pl}$, and $j_0^\text{Pl}$ fixed), we can get an expression for the perturbation of matter density, which we can associate to the local void. Replacing the Eq. \eqref{feins2} in the Eq. \eqref{feins1} we got

\begin{eqnarray}
c^2\nabla^2 \Phi=4\pi \mathbb{G}a^2\rho_m^{(1)}+3Ha^2(H\Phi+\dot{\Phi})\,.
\end{eqnarray}

Remember that $H=H(\tau)$ is obtained by the background cosmology Eq. \eqref{eq:friedmann}. Then

\begin{eqnarray}\label{rho1}
\rho_m^{(1)}=\frac{c^2\nabla^2 \Phi}{4\pi \mathbb{G}a^2}-\frac{3H^2\Phi-3H\dot{\Phi}}{4\pi \mathbb{G}}\,,
\end{eqnarray}

where the Laplacian in spherical coordinates is given by

\begin{eqnarray}
\nabla^2 f(r)=\frac{1}{r^2}\frac{d}{dr}\left(r^2\frac{d}{dr}f(r)\right)\,.
\end{eqnarray}

Replacing $\Phi(\tau,r)=F(\tau)G(r)$ in Eq. \eqref{rho1}, gives us the solution for $\rho_m^{(1)}$

\begin{eqnarray}
\rho_m^{(1)}(\tau,r)&=&\frac{3H^2(\tau)F(\tau)-3H(\tau)\dot{F}(\tau)}{4\pi \mathbb{G}}G(r)\nonumber\\&+&\frac{6c^2\nabla^2 G(r)F(\tau)}{4\pi \mathbb{G}a(\tau)^2}\,,
\end{eqnarray}

where $\tau=t-t_0$. Evaluating in the present ($\tau=0$)

\begin{eqnarray}
\rho_m^{(1)}(0,r)\!=\!\frac{6c^2\nabla^2 G(r)\Phi_0}{4\pi \mathbb{G}}\!-\!\frac{3{H_0^\text{Pl}}^2\Phi_0\!-\!3{H_0^\text{Pl}}f_1}{4\pi \mathbb{G}}G(r)\!\,.
\end{eqnarray}

\subsection{Profile of the void}

Until now, we have said anything about $g_4$, this is because in the procedure of section \ref{Sec4}, it is a free parameter. Nevertheless, the expansion of Eq. \eqref{G(r)} is only valid for the purpose of finding a series expansion of luminosity distance. For instance, if we want to describe a general profile, we need to expand $G(r)$ to higher orders:

\begin{equation}
    G(r)=\sum_{i=0}^n g_{2i}r^{2i}\,.
\end{equation}

In order to know every parameter $g_{2i}$ with respect to the parameters in the luminosity distance $q_0$, $j_0$, $s_0$ and higher order terms, those cosmological parameters have to be valid to redshift lower than $z=0.15$ to keep our procedure in agreement with SNe-Ia local measurements. However, at low redshift those cosmological parameters can not be measured because of a lack of precision in the fits. This inaccuracy emerges because of our regime in redshift, where the powers of $z$ are too smalls. Then, the deviation of these parameters tends to be large, see for instance \citep[][]{Capozziello_2019}. Hence, we can not find the parameters $g_{2i}$ as a function of $q_0$, $j_0$ and next order parameters. This does not imply that there is no density profile that can satisfy or not the Hubble tension. There is simply no data to assert or reject our hypothesis of a general function $G(r)$.\\
On the other hand, we can consider a simple Gaussian $G(r)=\exp{(-\alpha r^2)}$ as our hypothesis, then we only need $g_2$ to determine completely the profile, because in the expansion of $G(r)$ we have $\alpha=-g_2$. The main problem of this approach is that in the expansion shown in Eq. \eqref{G(r)} we cannot cut the series at any order because the complete Gaussian has to be valid for the scale of usual voids ($\sim 300-600$ Mpc), or in terms of the redshift: $z\sim 0.15$.

\section{Results}
We used numerical computations in order to constrain the local matter perturbations as a function of $q_0^\text{Lo}$ and $j_0^\text{Lo}$. In other words, we explored the parameter space of $q_0^\text{Lo}$ and $j_0^\text{Lo}$ solving the Eqs. \eqref{dLperturbed} and \eqref{dLStandard}. Where we have used Eq. \eqref{feins2} to obtain $v^N$. In order to keep a physical description for observational evidence we required a suitable void profile neglecting: complex solutions, a range of values of $\Phi_0$ outside the perturbative regime $\Phi_0>0.1$, overdensities described by negative potential $\Phi_0<0$ and $g_2>0$, and finally, a regime where the matter perturbation are not out of the range $\Omega^{(1)}_{m,0}/\Omega^{(0)}_{m,0}>-0.15$, and $\Omega^{(1)}_{m,0}/\Omega^{(0)}_{m,0}< -0.45$ (around the $30\%$ of the observed subdensity \cite{Bohringer}, \cite{Keenan_2013} and \cite{10.1093/ptep/ptz066}).

We emphasise that the results depend on the background, then the range of the parameters could change if the background cosmology changes (obtained from Planck). In this method, we have assumed that $\Lambda$CDM is the correct model. In this case, $q_0^\text{Pl}$ depends on the DE and matter parameters:

\begin{equation}
q_{0}^\text{Pl}={\frac {1}{2}}\Omega_{m,0}^\text{Pl}-\Omega_{\Lambda }^\text{Pl}.
\end{equation}

With the \cite{Planck2018} data, we fixed $q_0^\text{Pl} =-0.5275$. Regarding the second parameter, we assumed that $j_0^\text{Pl}=1$ as $\Lambda$CDM predicts.\\
With all of these considerations, we have determined that this approach does not yield any viable physical solutions.\vspace{0.3cm}\\ 
On the other hand, in $\Lambda(w)$CDM theory, it is usual to use the parametrization $\omega=\omega_0+\omega_a(1-a(t))$ \cite{Capozziello:2011tj}. This yields expressions for $q^\text{Pl}_0$ and $j^\text{Pl}_0$ as follows:
 \begin{eqnarray}
     q_0^\text{Pl}&=&\frac{1}{2}+\frac{3}{2}(1-\Omega^{(0)}_m)\omega_0\,,\\
     j_0^\text{Pl}&=&1+\frac{3}{2}(1-\Omega^{(0)}_m)[3\omega_0(1+\omega_0)+\omega_a]\,.
 \end{eqnarray}
 Using the Planck data \cite{Planck2018} for $\omega_0$ and $\omega_a$, we get $q_0^\text{Pl}=-0.48$, $j_0^\text{Pl}=0.58$. With the Hubble constant $H_0^\text{Pl}=68.31$ km Mpc$^{-1}$ s$^{-1}$, this approach does not offer any viable physical solution.

\section{Conclusions}\label{Conclusions}

We have explored the possibility of how local density perturbations could reconcile the Hubble tension between local and cosmological measurements of the Universe expansion. We perturbed the FLRW metric with a potential $\Phi$ which varies both in time and space. The temporal evolution is relevant because from redshift $z=0.15$ until now elapse approx 2 Gyrs. This is enough time for photons evolve with the temporal evolution of the potential due to the local void.\\
With this perturbation, we have computed the local luminosity distance, which depends on $H_0^\text{Lo}$, $q_0^\text{Lo}$ and $j_0^\text{Lo}$, and we compared it with the cosmological luminosity distance, that depends on $H_0^\text{Pl}$, $q_0^\text{Pl}$ and $j_0^\text{Pl}$. Due to our hypothesis about $\Phi(t,r)$, we could not find a region of $q_0^\text{Lo}$ and $j_0^\text{Lo}$ which allow us to explain the variation between $H_0^\text{Lo}$ and $H_0^\text{Pl}$.\\
This region of parameters was constrained such that the local density was $\Omega^{(1)}_{m,0}/\Omega^{(0)}_{m,0}=-0.3\pm 0.15$, in agreement with \cite{Bohringer}, \cite{Keenan_2013}, and \cite{10.1093/ptep/ptz066}. The range of values for $\Phi_0$ in our region is $0<\Phi_0 < 0.1$.\\
It is relevant to point out that we wanted that our whole region of $q_0^\text{Lo}$ and $j_0^\text{Lo}$ were in agreement with \cite{Riess_2016} results at redshift $z<0.15$; because the local luminosity distance is insensitive to higher-order terms, such as $q_0^\text{Lo}$ and $j_0^\text{Lo}$ that we explored in this work.\\
We want to emphasize that our scenario is different from previous studies where the inhomogeneity is described as a contribution in the spatial curvature of the FLRW metric through the LTB metric. In these studies, \cite{PhysRevD.103.123539,Kenworthy_2019} the conclusion strengths that a local void cannot save the tension on the Hubble constant.\\
On the other hand, the idea that local underdensities can be the origin of the Hubble discrepancy has also been studied using non-linear evolution \citep[][]{LOMBRISER2020135303,10.1093/mnras/stx1967}, where the underdensities needed are extreme ($\delta_m \approx -0.8 $ and $-0.5$ respectively).  \cite{Bohringer} studied the local underdensity through the CLASSIX galaxy cluster survey. They found that part of the $H_0$ discrepancy can be explained with the local underdensity which was determined as a void of $\delta_m= -30\pm 15$\% ($-20\pm 10$\%) in a region with a radius of about $100$ ($\sim 140$) Mpc.\\
 Conversely, we worked in the Newtonian gauge for the standard perturbed FLRW metric (the same perturbed metric as in the CMB \cite{weinberg_book, Piattella_book}) and not the LTB metric. Our finding showed that the Hubble tension cannot be solved for this range of underdensities (for this very simplified model). Also we explored the cosmological parameters $H_0^\text{Lo}$, $q_0^\text{Lo}$ and $j_0^\text{Lo}$ found in \cite{Capozziello_2019} and \cite{Capozziello:2011tj}, but, there is also no possible solution in our approach based on $\Lambda(\omega)$CDM model.

\section*{Acknowledgements}
The author C.R. is supported by FONDECYT grant 3220876.



\bibliography{apssamp}
\appendix

\section{Coefficients $f_2$, $f_3$ and $f_4$}\label{f2f3f4}
The obtained coefficients are:
\begin{eqnarray}
f_2&=&\frac{1}{2}\left(-\Phi_0{H_0^\text{Pl}}^2-4f_1H_0^\text{Pl}+2\Phi_0{H_0^\text{Pl}}^2q_0^\text{Pl}\right)\;\label{eq:f_2},\\
f_3&=&\frac{1}{6}\left(6\Phi_0{H_0^\text{Pl}}^3+19f_1{H_0^\text{Pl}}^2-2\Phi_0{H_0^\text{Pl}}^3j_0^\text{Pl}\right.\nonumber\\&-&\left.8\Phi_0{H_0^\text{Pl}}^3q_0^\text{Pl}+6f_1{H_0^\text{Pl}}^2q_0^\text{Pl}\right)\;,\label{eq:f_3}\\
f_4&=& \frac{1}{24} \left(10 \Phi_0{H_0^\text{Pl}}^4
   j_0^\text{Pl}+20 \Phi_0 {H_0^\text{Pl}}^4
   {q_0^\text{Pl}}^2\right.\nonumber\\&+&\left.30 \Phi_0 {H_0^\text{Pl}}^4 q_0^\text{Pl}-37 \Phi_0
   {H_0^\text{Pl}}^4-8 f_1 {H_0^\text{Pl}}^3 j_0^\text{Pl}\right.\nonumber\\&-&\left.76
   f_1 {H_0^\text{Pl}}^3 q_0^\text{Pl}-108 f_1
   {H_0^\text{Pl}}^3\right)\;, \label{eq:f_4}
\end{eqnarray}

where $\Phi_0$ and $f_1$ as free parameters, because \eqref{eq:F_diff_eq} is a second order equation.\\

\section{Coefficients $\tau_i$}\label{coefficients}
The solutions for $\tau_i$ are
\begin{eqnarray}
\tau_1&=&-1+2f_0\\
\tau_2&=&\frac{{H_0^\text{Pl}}}{2}-2 f_0
   {H_0^\text{Pl}}-f_1\\
 \tau_3&=&\frac{1}{3}  \left(2 c^2 f_0 g_2-f_0
   {H_0^\text{Pl}}^2 q^{\text{Pl}}+2 f_0
   {H_0^\text{Pl}}^2\right)\nonumber\\&+&\frac{1}{6} \left({H_0^\text{Pl}}^2
   q_0^\text{Pl}-{H_0^\text{Pl}}^2\right)\\
 \tau_4&=&\frac{1}{12}  \left(-8 c^2 f_0 g_2
   {H_0^\text{Pl}}-6 c^2 f_1 g_2-2 f_0 {H_0^\text{Pl}}^3
   j_0^\text{Pl}\right.\nonumber\\&+&\left.10 f_0 {H_0^\text{Pl}}^3 q_0^\text{Pl}-3
   f_0 {H_0^\text{Pl}}^3+f_1 {H_0^\text{Pl}}^2 q_0^\text{Pl}-2
   f_1 {H_0^\text{Pl}}^2\right)\nonumber\\&+&\frac{1}{24}
   \left({H_0^\text{Pl}}^3 j_0^\text{Pl}-4 {H_0^\text{Pl}}^3
   q_0^\text{Pl}+{H_0^\text{Pl}}^3\right)\\
 \tau_5&=&\frac{1}{60}  \left(24 c^4 f_0 g_4+4 c^2 f_0
   g_2 {H_0^\text{Pl}}^2 q_0^\text{Pl}+8 c^2 f_0 g_2
   {H_0^\text{Pl}}^2\right.\nonumber\\&-&\left.6 c^2 f_1 g_2 {H_0^\text{Pl}}+23 f_0
   {H_0^\text{Pl}}^4 j_0^\text{Pl}+12 f_0 {H_0^\text{Pl}}^4
   {q_0^\text{Pl}}^2\right.\nonumber\\&-&\left.35 f_0 {H_0^\text{Pl}}^4 q_0^\text{Pl}+2 f_0
   {H_0^\text{Pl}}^4+3 f_1 {H_0^\text{Pl}}^3 j_0^\text{Pl}\right.\nonumber\\&-&\left.13
   f_1 {H_0^\text{Pl}}^3 q_0^\text{Pl}-f_1
   {H_0^\text{Pl}}^3\right)+\frac{1}{120} \left(-7
   {H_0^\text{Pl}}^4 j_0^\text{Pl}\right.\nonumber\\&-&\left.4 {H_0^\text{Pl}}^4
   {q_0^\text{Pl}}^2+11 {H_0^\text{Pl}}^4
   q_0^\text{Pl}-{H_0^\text{Pl}}^4\right)
\end{eqnarray}
\section{$r(z)$}\label{r(z)}
In order to obtain $r(z)$, the luminosity distance shown in Eq. \eqref{luminositydistance} has to be integrated. We get an expression for $r(z)$ using Eqs. \eqref{zunperturbed}, \eqref{zperturbed},\eqref{eq:tau} and the coefficients $f_i$ from Eqs. \eqref{eq:f_2}, \eqref{eq:f_3} and \eqref{eq:f_4}:
 
\begin{eqnarray}
r(z)&=&\frac{cz}{H_0^\text{Pl}}\left[\left(1+2\Phi_0+\frac{ 
   f_1}{{H_0^\text{Pl}}}-\frac{4c^2\Phi_0g_2}{3{H_0^{\text{Pl}}}^2}-\frac{4c^2f_1g_2}{3{H_0^\text{Pl}}^3}\right)\right.\nonumber\\
   &+&\left.z \left(\frac{ c^2 \Phi_0 g_2}{3{H_0^\text{Pl}}^2}-\frac{3
    f_1 q_0^\text{Pl}}{2{H_0^\text{Pl}}}+\frac{6c^2\Phi_0g_2q_0^{\text{Pl}}}{{H_0^\text{Pl}}^2}-\frac{2c^2f_1g_2}{{H_0^\text{Pl}}^3}\right.\right.\nonumber\\&+&\left.\frac{14c^2f_1g_2q_0^\text{Pl}}{3{H_0^\text{Pl}}^3}-2 \Phi_0
   q_0^\text{Pl}
   -\frac{\Phi_0}{2}-\frac{ q_0^\text{Pl}}{2}-\frac{1}{2 }\right)\nonumber\\&+&z^2
   \left(-\frac{ 13c^2 f_1 g_2}{9{H_0^\text{Pl}}^3}-\frac{38
   c^2 \Phi_0 g_2 q_0^\text{Pl}}{9{H_0^\text{Pl}}^2}-\frac{7
   c^2 \Phi_0 g_2}{9{H_0^\text{Pl}}^2}-\frac{2 f_1
   j_0^\text{Pl}}{3 {H_0^\text{Pl}}}\right.\nonumber\\
   &+&\left.\frac{26c^2\Phi_0g_2j_0^\text{Pl}}{9{H_0^\text{Pl}}^2}-\frac{18c^2\Phi_0g_2{q_0^\text{Pl}}^2}{{H_0^\text{Pl}}^2}+\frac{20c^2f_1g_2j_0^\text{Pl}}{9{H_0^\text{Pl}}^3}\right.\nonumber\\
   &-&\left.\frac{38c^2f_1g_2{q_0^\text{Pl}}^2}{3{H_0^\text{Pl}}^3}+\frac{16c^2f_1g_2q_0^\text{Pl}}{9{H_0^\text{Pl}}^3}-\frac{8c^4g_4}{3{H_0^\text{Pl}}^4}-\frac{8c^4f_1g_4}{3{H_0^\text{Pl}}^5}\right.\nonumber\\
   &+&\left.\frac{5  f_1
   {q_0^\text{Pl}}^2}{2{H_0^\text{Pl}}}+\frac{ f_1}{6
   {H_0^\text{Pl}}}-\frac{2  \Phi_0 j_0^\text{Pl}}{3}
   +3  \Phi_0
   {q_0^\text{Pl}}^2+\frac{\Phi_0}{3}\right.\nonumber\\&+&\left.\frac{f_1q_0^\text{Pl}}{{H_0^\text{Pl}}}+\left.\frac{5  \Phi_0
   q_0^\text{Pl}}{3}-\frac{
   j_0^\text{Pl}}{6}+\frac{
   {q_0^\text{Pl}}^2}{2}+\frac{2 
   q_0^\text{Pl}}{3}+\frac{1}{3}\right)\right]\nonumber\\&+&O(z)^4\,.
\end{eqnarray}
\end{document}